\documentclass{article}
\addtocounter{equation}{0}
\begin{document}
\title{Energy quantization for matter orbiting black hole and Hawking radiation}         % Enter your title between curly braces
\author{E. \v{S}im\'{a}nek \footnote {Electronic address: simanek@ucr.edu}\\Department of Physics, University of California, Riverside, CA 92521}      % Enter your name between curly braces
\date{}          % Enter your date or \today between curly braces
\maketitle

\begin{abstract}

The energy of a test particle orbiting a Schwarzschild black hole is quantized owing to the quantization of the angular momentum. For smallest stable circular orbit, the excitation energy is found to resemble closely the expression for the temperature of the Hawking radiation. This result is consistent with the Unruh effect for orbiting test particle.  The predicted energy quantization might be observable by studies of the red-shifted 21-cm line of neutral hydrogen orbiting a primordial  black hole with mass of the order of that of Earth.

\end{abstract}

PACS number(s):  04.70.Bw, 04.70.Dy, 95.30.Sf
\\[10pt]

\section{Introduction}

	Long ago, Dirac showed that the existence of magnetic monopoles implies quantization of electric charge [1].  In 1985, Zee [2] proposed a gravitational analog of Dirac quantization condition.  This analog is based on the theory of gravitoelectromagnetism (GEM) [3].  In ordinary GEM, which is the analog of Maxwell theory it is the mass of the test particle and the gravitational acceleration that play the role of the electric charge and the electric field, respectively.  The GEM magnetic field is given by the GEM analog of the Biot-Savart law in which the source is the matter current density.  This field is a divergenceless quantity everywhere.  Following Dirac, Zee [2] considers the possibility that a GEM analog of the magnetic monopole, a "gravitipole", exists.  Consequently, the magnetic field of the gavitipole satisfies the equation

\begin{equation}\label{1}
\vec{\nabla}. \vec{B} =\gamma \delta^{3} (\vec{x})
\end{equation}
Using the Gauss law, we can integrate Eq. (1) to observe that the constant $\gamma$ is equal to the total magnetic flux $\Phi$ emanating from the gravitipole.  The quantization of the mass of the text particle is obtained by expanding the action $S$ of the test particle in the gravitational field.  Zee [2] focuses on the term representing the interaction of the test particle with the vector potential $\vec {\zeta}$ of the gravitipole

\begin{equation}\label{2}
S_{int} = - m \int d \vec{x} . \vec{\zeta} = - m \Phi
\end{equation}
where $m$ is the rest mass of the test particle. As the test particle is moved along a closed loop around the gravitipole, its wave function acquires a phase $S_{int} / \hbar = - m \gamma / \hbar$.  Single-valuedness implies that this phase be equal to $2 \pi n$ where $n$ is an integer.  This yields the quantization condition [2]

\begin{equation}\label{3}
m = \frac {2 \pi \hbar n} {\gamma}
\end{equation}
Owing to the hypothetical nature of the gravitipole, the constant $\gamma$ is not known.  Only the upper bound of the energy splitting can be estimated from the absence of the quantization effects on energy level splittings in atoms and molecules [2].

The present effort was inspired by the paper of Zee [2].  We consider a quantization of the energy that is due to the gravitational field of a black hole.  In distinction from Ref. [2], it is not the rest mass of the test particle that becomes quantized.  Rather, it is the radius of the particle orbit that is quantized as a result of quantizing the angular momentum. For a test particle, moving along a circular path around the black hole, the quantization is due to the fact that the azimuthal angle is a periodic function of time.  Since the canonical momentum that is conjugate to this angle is the angular momentum, the corresponding action variable is the angular momentum times $2 \pi$.  Applying the postulate of the old quantum theory we set the action variable equal to $n h$ where $n$ is an integer and $h$ is the Planck constant [4,5].  In this way, the energy quantization is linked to the quantization of the angular momentum of the test particle orbiting a black hole.

In Sec. 2, we consider a Schwarzschild black hole and derive the Lagrangian and the canonical momenta for the test particle. Using these momenta, we obtain in Sec. 3 the effective potential and the radial equation of motion of the time-like geodesics.  From the effective potential, we determine the condition that the angular momentum must satisfy in order that the geodesics are stable circular orbits.  In Sec. 4, we derive the quantization condition for the energy of the test particle.  We find that the mass quantum is a function of the mass of the black hole and the radius of the orbit.  For the smallest stable orbit, the formula for the excitation energy resembles closely the expression for the temperature of the Hawking thermal radiation [6].  In Sec. 4, we attempt to clarify this similarity by comparing the period of the orbital motion with the period of the Euclidean time.  Sec. 5 is devoted to estimates of the energy quantization in various astronomical settings. The relation of the excitation energy to Hawking temperature is discussed in the Appendix in terms of the Unruh effect [22].

\section{Lagrangian and canonical momenta of test particle}

	We consider a test particle of mass $m$ orbiting a Scharzschild black hole. We confine ourselves to circular geodesics in the equatorial plane.  Then the metric takes the form [7,8]

\begin{equation}\label{4}
ds^{2} = -(1- \frac{r_{S}}{r})c^{2} dt^{2} + (1 - \frac{r_S}{r})^{-1} dr^{2} + r^{2} d \phi^{2}
\end{equation}	
where $r_{S} = 2 MG/c^{2}$ is the Schwarzschild radius for a black hole of mass $M$.  We assume that $M$ is much larger than the Planck mass, ensuring that $r_S$ is much larger than the Compton radius, $r_C = \hbar/Mc$, so that quantum fluctuations of the black hole (zitterbewegung) can be disregarded [9].

The Lagrangian of the test particle can be expressed in terms of the metric components $g_{ij}$ as follows [7,8]

\begin{equation}\label{5}
\pounds = - \frac{m}{2} g_{ij} \dot{x}^{i} \dot{x}^{j} = - \frac {m}{2} (g_{00} c^{2} \dot {t}^{2} + g _{rr} \dot{r}^{2} + g _{\phi \phi} \dot{\phi}^{2})
\end{equation}
where we adopt the notation 

\begin{equation}\label{6}
\dot{x} ^{0} = c \dot {t} = c \frac{dt}{d\tau}, \dot{x}^{r} = \frac{dr}{d\tau}, \dot{x}^{\phi} = \frac {d \phi} {d \tau}
\end{equation}
Since we are considering time-like geodesics, we choose $\tau$ to be the proper time.

The canonical momenta, $p_{\alpha} = \partial \pounds/\partial \dot{x}^{\alpha}$, are obtained with the use of Eqs. (4) and (5) yielding

\begin{equation}\label{7}
p_{t} = \frac{\partial \pounds} {\partial \dot{t}} = - m c^{2} g_{00} \dot{t} = m c ^{2} (1 - \frac{r_{S}}{r}) \dot{t} = E
\end{equation}

\begin{equation}\label{8}
p_{\phi} = \frac{\partial \pounds} {\partial \dot{\phi}} = - m g _{\phi \phi} \dot {\phi} = - m r^{2} \dot{\phi} =- L
\end{equation}

\begin{equation}\label{9}
p_{r} = \frac{\partial \pounds} {\partial \dot{r}} = - mg_{rr} \dot{r} =- m (1 - \frac{r_{S}}{r})^{-1} \dot{r}
\end{equation}
where $E$ and $L$ are the total energy and angular momentum of the test particle.

\section {Effective potential for radial motion}

	The equation for radial motion of a geodesic is obtained from the relation

\begin{equation}\label{10}
g^{00} p^{2}_{0} + g^{rr} p^{2}_{r} + g^{\phi\phi} p^{2}_{\phi} + m^{2} c^{2} = 0
\end{equation}
where the first three terms describe the magnitude of the energy-momentum four-vector $(p_{0}= E/c , \vec{p})$.
Introducing into Eq. (10) the quantities $g^{ij}$ from Eq. (4), and the canonical momenta from Eqs. (7-9), we obtain [8]

\begin{equation}\label{11}
- \frac{E^{2}}{c^{2}} (1 - \frac{r_{S}}{r})^{-1} + m^{2} (1 - \frac{r_{S}}{r})^{-1} \dot{r}^{2}+ \frac{L^{2}}{r^{2}} + m^{2} c^{2} = 0
\end{equation}

Following the notation of Ref. [7], we introduce the energy and momentum of the test particle per unit rest mass

\begin{equation}\label{12}
\tilde{E} = \frac{E}{m} , \tilde{L} = \frac{L}{m}
\end{equation}
Introducing these definitions into Eq. (11), we obtain

\begin{equation}\label{13}
m^{2}c^{2} \Big[- \frac{\tilde{E}^{2}}{c^{4}} (1 - \frac{r_{S}}{r})^{-1} + \frac{\dot {r}^{2}}{c^{2}} (1 - \frac{r_{S}}{r})^{-1} + \frac{\tilde{L}^{2}}{c^{2}r^{2}} + 1\Big] = 0
\end{equation}
From this equation, we obtain the equation for radial motion in the form

\begin{equation}\label{14}
\frac{\dot{r}^{2}}{2} =  \frac{\tilde{E}^{2}}{2 c^{2}} - \frac {1}{2} \Big(\frac{\tilde{L}^{2}}{r^{2}} + c^{2}\Big) \Big(1 - \frac{r_{S}}{r}\Big)
\end{equation}
Letting $G=c=1$, this equation takes the form of Eq. (6.3.14) of Ref. [8].  The effective potential for the radial motion is obtained from Eq. (14) in the form

\begin{equation}\label{15}
V_{eff} = \frac{1}{2} \Big(1- \frac{r_{S}}{r}\Big) \Big(\frac{\tilde{L}^{2}}{r^{2}} + c^{2} \Big)
\end{equation}
The extrema of this potential are found from the equation

\begin{equation}\label{16}
\frac{\partial V_{eff}}{\partial r} = \frac{c^{2}r_{S}}{2 r^{4}} \Big(r^{2} - \frac{2 \tilde{L}^{2}}{c^{2} r_{S}} r + \frac{3 \tilde{L}^{2}} {c^{2}}  \Big) = 0
\end{equation}
The roots of this equation are

\begin{equation}\label{17}
R_{\pm} = \frac{\tilde{L}^{2}}{c^{2}r_{S}} \pm \Big[\Big(\frac{\tilde{L}^{2}}{c^{2}r_{S}}  \Big)^{2} - \frac{3 \tilde{L}^{2}} {c^{2}}  \Big]^{\frac{1}{2}}
\end{equation}
We see that the roots are real only if $\tilde{L}^{2} > 3 c^{2} r^{2}_{S}$. In this case, $R_{ +}$ is a true minimum of the effective potential.  This can be verified by computing, using Eq. (15), the second derivative of the effective potential which turns out positive at $r = R_{ +}$.

In what follows, we consider only stable circular orbits for which $\tilde{L}^{2} > 3 c^{2} r ^{2}_{S}$ implying, according to Eq. (17), that the radius of the orbit $R$, must be larger than $3 r_{S}$.

\section{Energy quantization and Hawking temperature}

	As pointed out in Sec. 1, the quantization of energy is closely related to the quantization of the angular momentum of the orbiting test particle.  This stems from the fact that the angular variable $\phi$ is a periodic function of time so that we may apply the prescription of Wilson [4] and Sommerfeld [5]

\begin{equation}\label{18}
J_{\phi} = \int^{2 \pi}_{0} L d \phi = n h
\end{equation}
where $J_{\phi}$ denotes the action variable for one period of motion, and $L$ is the canonical momentum conjugate to the angular variable (see Eq. (8)).

Since $L$ is a constant of motion, Eq. (18) yields the quantization condition for the angular momentum

\begin{equation}\label{19}
L = m \tilde{L} = n \hbar
\end{equation}
Equation (19) implies, via Eq. (17), a quantization of the radius $R$. For the smallest stable orbit, $R_{0} = 3 r_{S}$, the square root on the right hand side of Eq. (17) vanishes so that 

\begin{equation}\label{20}
R_{0} = \frac{\tilde{L}^{2}_{0}}{c^{2}r_{S}} = \frac {\hbar^{2} n^{2}_{0}} {m^{2} c^{2} r_{S}}
\end{equation}
yielding

\begin{equation}\label{21}
n_{0} = \frac{3^{\frac{1}{2}} r_{S}}{r_{C}}
\end{equation}
where $r_{C} = \frac{\hbar}{mc}$ is the Compton radius of the test particle.  The next higher orbit corresponding to $n_{1} = n_{0} + 1$, has radius $R_{1}$ given by Eq. (17) in the form 

\begin{equation}\label{22}
R_{1} = (n_{0} + 1)^{2} \frac{r^{2}_{c}}{r_{S}} \Big \lbrace 1 + \Big [ 1 - \frac{3 r^{2}_{S}}{r^{2}_{C}(n_{0}+ 1)^{2}} \Big ] ^ {\frac{1}{2}} \Big \rbrace
\end{equation}

For the sake of simplicity, we assume that $r_{S} \gg r_{C}$. Eq. (21) then implies that $n_{0} \gg 1$ and $(n_{0} + 1)^{2} \approx n^{2}_{0} (1 + \frac{2}{n_{0}})$ Expanding Eq. (22) in a small parameter $2/n_{0}$, we obtain

\begin{equation}\label{23}
R_{1} \approx R_{0} \Big [ 1 + \Big (\frac{2}{n_{0}} \Big ) ^{\frac{1}{2}} \Big ]
\end{equation}

The true energy $E = m \tilde {E}$ is obtained from Eq. (14), yielding

\begin{equation}\label{24}
\tilde {E}^{2} = c ^{2} \Big ( \frac{\tilde{L}^{2}} {R^{2}} + c^{2} \Big )
 \Big ( 1 - \frac{r_{S}}{R} \Big )
\end{equation}
From Eq. (16), we obtain

\begin{equation}\label{25}
\frac{\tilde{L}^{2}}{R^{2}} = \frac {c^{2}} {2 R/r_{S}-3}
\end{equation}
Substituting Eq. (25) into (24) gives the true energy as a function of $R$

\begin{equation}\label{26}
E = m c^{2} \frac{1 - r_{S}/R} {(1- \frac{3}{2} r_{S}/R)^{\frac{1}{2}}}
\end{equation}
For $R_{0}= 3 r_{S}$, we obtain from this equation

\begin{equation}\label{27}
E_{0} = \frac{2 mc^{2}}{3 (0.5)^{\frac{1}{2}}} = 0.94 mc^{2}
\end{equation}
For $R_{1}$, given by Eq. (23), the true energy is given by

\begin{equation}\label{28}
E_{1} \approx E_{0} \frac{1 + \frac{1}{2}(x - x^{2})} {(1 + x- x^{2})^{\frac{1}{2}}}
\end{equation}
where $x = (2/n_{0})^{\frac{1}{2}}$.  Expanding in small parameter $x$, we obtain from Eq. (28) to order $x^{2}$

\begin{equation}\label{29}
E_{1} \approx E_{0} (1 + \frac{1}{4 n_{0}})
\end{equation}
Then the excitation energy $E_{1} - E_{0} = \delta E_{10}$ is obtained from Eqs. (21), (27) and (28) as follows

\begin{equation}\label{30}
\delta E_{10} \approx \frac{0.94}{4} \frac{mc^{2}}{n_{0}} \approx 0.068 \frac{\hbar c^{3}}{MG}
\end{equation}
This formula shows remarkable resemblance to the expression for the temperature of the Hawking thermal radiation [6]

\begin{equation}\label{31}
k_{B} T_{H} = \frac{\hbar c^{3}}{8 \pi MG}
\end{equation}

This similarity can be understood by examining the period of the angular variable $\phi (\tau)$ for the particle encircling the black hole, and comparing it with the period of the Euclidean time $\tau _{E} = it$. For a circular orbit of radius $R$, the "orbital" period is given by

\begin{equation}\label{32}
\tau _{o} = \frac{2 \pi R}{v_{\phi}} = \frac{2 \pi R ^{2}}{\tilde {L}} \dot{t}
\end{equation}
where we used Eq. (8) to write $v_{\phi} = \frac{R \dot{\phi}}{\dot{t}} = \frac{\tilde{L}}{R \dot{t}}$.  Using Eq. (25), we can express $\tilde{L}/R$ in terms of $R$, and obtain from Eq. (32) 

\begin{equation}\label{33}
\tau_{o} = \frac{2 \pi R}{c} \Big(\frac{2R}{r_{S}} - 3 \Big)^{\frac{1}{2}} \Big( 1 - \frac{3}{2} r_{S}/R \Big ) ^{-\frac{1}{2}}
\end{equation}
where we used Eq. (A3) for $\dot{t}$. Substituting the radius of the smallest stable orbit, $R = 3 r_{S}$, Eq. (33) yields

\begin{equation}\label{34}
\tau_{o} = 6 ^{\frac{3}{2}} \pi \frac{r_{S}}{c}
\end{equation}

Next, we consider the period of the Euclidean time $\tau _{E}$.  Letting $\tau_{E} = it$ in Eq. (4), we obtain the Euclideanized metric.  Following Ref. [10], we require this metric to be regular at the horizon.  Thus, we define a small quantity $\rho$ by

\begin{equation}\label{35}
r = r_{S} (1 + \rho^{2})
\end{equation}
and expand the metric about $\rho = 0$.  In this way, we obtain in the vicinity of the horizon

\begin{equation}\label{36}
d s^{2} = 4 r^{2}_{S} \Big ( d \rho^{2} + \frac{\rho^{2} c ^{2}} {4 r^{2}_{S}} d \tau^{2}_{E} \Big) + r^{2}_{S} d  \phi^{2}
\end{equation}

The first two terms of Eq. (36) correspond to a metric $d \rho^{2} + \rho^{2} d \tilde{\phi}^{2}$ in polar coordinates $\rho, \tilde {\phi}$ in flat plane.  Since $\tilde {\phi}$ has period $2 \pi$, Eq. (36) implies that the period of $\tau_{E}$ is 

\begin{equation}\label{37}
\tau_{E} = \frac{4 \pi}{c} r _{S}
\end{equation}
Except for the dimensionless prefactor, the similarity to the orbital period (34) is apparent.

The function $\phi (\tau_{E})$ also exhibits a period $\hbar \beta$ which is obtained by considering the partition function $Z = T_{r} (e ^{-\beta H})$ where $\beta = (k_{B} T)^{-1}$. For particle on a circle, $Z$ can be expressed as a path integral with topological constraint making $\phi = 0$ and $\phi = 2 \pi$ indistinguishable.  $Z$ is then given as a sum of path integrals involving paths with various winding numbers $l$[11].  These paths satisfy the condition $\phi (\tau_{E} = \hbar \beta) = \phi (\tau_{E} = 0) + 2 \pi l$.  In view of the topological constraint, path differing by $2 \pi l$ are indistinguishable, implying that the period of $\phi (\tau_{E})$ is $\tau_{E} = \hbar \beta$.  Combining this result with Eq. (37), we obtain 

\begin{equation}\label{38}
k _{B}T_{H} = \frac{\hbar c} {4 \pi r_{S}}
\end{equation}
which agrees with Eq. (31).

The fact that $\delta E$ and $k _{B} T_{H}$ are quantities of the same order of magnitude suggests that interaction of photons with matter may play role in establishing the thermal equilibrium of the black body radiation [12].  For matter with quantized energy the transitions $n \rightarrow n \pm 1$ correspond to absorption and emission of photons by matter.  These effects should be particularly strong when $\delta E = \hbar \omega_{m}$ = 2.8 k$_{B}$ $T_{H}$, where $\omega_{m}$ is the frequency at which the Planck formula for spectral distribution of black body radiation is maximum [12]. Using Eq. (31), we obtain $\hbar \omega_{m} \approx 0.1 \frac{\hbar c^{3}}{MG}$, which is in reasonable agreement with $\delta E_{10}$ given by Eq. (30).  This implies that matter orbiting the black hole may be instrumental in establishing the thermal equilibrium of the radiation due to Hawking evaporation. There is another more direct way to demonstrate the relation between the dynamics of an orbiting particle and the Hawking radiation.  This is the Unruh effect discussed in the Appendix.

Now, we consider large orbits such that $R \gg r_{S}$.

From Eq. (17), we have

\begin{equation}\label{39}
R = \frac{\tilde {L} ^{2}} {c^{2} r_{S}} \Big [1 + \Big ( 1 - \frac{3 c^{2}r^{2}_{S}}{\tilde{L}^{2}}\Big)^{\frac{1}{2}} \Big ]
\end{equation}

Using the condition of mechanical stability, $\tilde {L}^{2} = M GR$, the second term in the parenthesis can be written as

\begin{equation}\label{40}
\frac{3 c^{2}r^{2}_{S}}{\tilde{L}^{2}} = \frac{6 r_{S}} {R}\ll 1
\end{equation}
Thus, to lowest order in $r_{S}/R$, the radius of the orbit becomes

\begin{equation}\label{41}
R \approx \frac{2 \tilde{L}^{2}} {c^{2} r _{S}}
\end{equation}

To obtain the energy as a function of $R$, we start from Eq. (26).  Expanding this equation in a small parameter $r_{S}/R$, we obtain to lowest order in $r_{S}/R$

\begin{equation}\label{42}
E \approx m c^{2} \Big (1 - \frac {r_{S}}{4 R} \Big )
\end{equation}
Introducing the radius $R$ from Eq. (41), Eq. (42) yields

\begin{equation}\label{43}
E \approx m c ^{2} \Big ( 1 - \frac{r^{2}_{S} c ^{2}} { 8 \tilde {L}^{2}} \Big )
\end{equation}
The quantized energy $E _{n}$ follows from Eq. (43) by replacing $\tilde {L}$ by quantized angular momentum $\tilde {L}_{n} = n \hbar/m$.  In this way, we obtain

\begin{equation}\label{44}
E_{n} \approx m c^{2} \Big ( 1 - \frac{r^{2}_{S}} {8 n^{2}r^{2}_{C}} \Big )
\end{equation}
The dependence of $E_{n}$ on the quantum number $n$ is reminiscent of Bohr's quantization of the total energy of an atomic electron as expected for a Kepler problem. Using this result, we obtain the energy splitting

\begin{equation}\label{45}
\delta E_{n} = E _{n + 1} - E_{n} \approx \frac{m c^{2} r ^{2}_{S}} {8 r ^{2}_{C}} \Big [ \frac{1}{n^{2}} - \frac{1} {(n + 1)^{2}} \Big ]
\end{equation}
For large values of $n$, the bracket can be replaced by $2/n^{3}$ so that

\begin{equation}\label{46}
\delta E_{n} \approx \frac{m^{3}r^{2}_{S} c^{4}} {4 \hbar ^{2} n ^{3}} = \frac{\hbar M^{2} G ^{2}}{\tilde{L}^{3}}
\end{equation}
where $n$ has been eliminated using Eq. (19).  Invoking the mechanical stability condition $\tilde {L}^{2} = MGR$, we obtain from Eq. (46)

\begin{equation}\label{47}
\delta E_{n} \approx \frac{\hbar (MG)^{\frac{1}{2}}}{R^{\frac{3}{2}}} \approx \frac{\hbar v} {R}
\end{equation}
where the second equality stems from the relation $v^{2} = MG/R$ which is equivalent to the mechanical stability condition as seen by writing $\tilde {L} = v R$. Incindentally, this result agrees with the Heisenberg uncertainty principle for energy and time.

However, this Newtonian result becomes modified outside the galaxy core where $v (R)$ tends to a constant value $v_{o}$.  This has been discovered by measuring the Doppler shift in the 21 cm-line of neutral hydrogen in the galaxy disc [13].  This behavior can be attributed to the fact that there is a mass distribution

\begin{equation}\label{48}
M (R) \approx \frac{v^{2}_{o} R} {G}
\end{equation}

where $M (R)$ is the mass within the radius $R$.  Hence, the corresponding mass density goes as $\rho (r) \propto r^{-2}$ indicating that there is a "hallo" of dark matter which stretches well beyond the visible part of the galaxy [13]

\section { Observational considerations}

In this section, we want to distinguish between "ordinary" black holes (OBHs) and primordial black holes (PBHs).  The OBHs are formed by collapse of massive stars (perhaps $10-10^{10}$ times heavier than our Sun).  The much lighter PBHs may have been formed by collapse of the overdense regions shortly after the Big Bang [8].  They are an important potential dark matter candidate [14, 15].

We first consider the supermassive black hole at the center of our own Milky Way galaxy.  Its mass, obtained by carefull astrometric measurements [16], is $M \approx 4 \times 10 ^{6} M_{\odot} $ where $ M _{\odot} = 1.989 \times 10^{33 } g$ is the solar mass. The corresponding Schwarzschild radius is $r_{S} \approx 1.2 \times 10^{12}$ cm. For the smallest stable orbit of radius $R = 3 r_{S}$, Eq. (30) yields $\delta E _{10} \approx$ 2.1 $\times$ 10$^{-18}$ eV. Owing to the large mass of the black hole this quantity is extremely small.  Clearly, the OBHs are not expected to produce a significant increase of $\delta E_{10}$ since their mass scale is of order 10 $M_{\odot}$ or more.

On the other hand, the PBHs can be produced at a mass scale much smaller than $M_{\odot}$.  For example, a black hole of mass equal to that of the Earth $\approx$ 6 $\times$ 10$^{27}$g can be produced at time $\tau \approx$ 10 $^{-11}$ sec after the Big Bang [8].  With the corresponding Schwarzschild radius $r_{S} \approx$ 0.9cm, the formula (30) yields $\delta E _{10} \approx$ 2.8 $\times$ 10 $^{-6}$ eV.

Let us assume that a test particle with mass of the hydrogen atom is orbiting this black hole at the radius $3 r_{S}$. The corresponding Compton radius $r_{C} = \frac {\hbar}{mc}$ for this test particle is $r_{C} \approx $ 2 $\times$ 10 $^{-14}$ cm.  The relevance of $r_{C}$ comes from the fact it gives the distance over which a relativistic particle of mass $m$ is spread out about its position [9, 17].  For a test particle moving in the gravitational field of the black hole, the potential seen by the test particle becomes effectively averaged over a region of size $r_{C}$.

If $r_{C} \ll r_{S}$, the test particle behaves as a classical point mass moving in the potential given by Eq. (4).  In this limit, the present semiclassical quantization approach is valid.

On the other hand, as $r_{C}$ grows and approaches the value of $r_{S}$, the present approach ceases to be valid and should be replaced by solving the complete quantum mechanical problem of Klein-Gordon scalar field in the space-time of the black hole [8].

The criterion of validity of the present approach, $r_{S} \gg r _{C}$, can be written as follows

\begin{equation}\label{49}
2 m M \gg M^{2}_{P}
\end{equation}
where $m$ is the mass of the test particle, $M$ is the black hole mass, and $M_{P}$ is the Planck mass $M_{P} = (\hbar c /G)^{\frac{1}{2}} \approx $ 1.2 $\times$ 10 $^{19}$ GeV/c$^{2} \approx$ 2.2 $\times $ 10 $^{-5}$ g.

For the above considered case of hydrogen atom in the gravitational field of a black hole with mass equal to that of the Earth, the inequality (49) is well satisfied.  However, a problem arises for PBHs that are lighter than about $M \approx$ 10$^{15}$g. The importance of these black holes derives from the fact that they would shed all their mass by Hawking evaporation by now [18].  Inserting $M \approx$ 10 $^{15}$ g into Eq. (49), we see that the rest mass of the test particle should be much larger than about 2.4 $\times$ 10 $^{-25}$ g.  This implies that our approach is marginally valid for $m$ equal to that the hydrogen atom $ \approx$ 1.67 $\times$ 10 $^{-24}$ g, but fails for particles such as mesons and especially for neutrinos. For such particles, the energy quantization and its role in equilibration of the emitted black body radiation, discussed in the previous section, may not be realized.  Obviously, as $M$ goes well below 10 $^{15}$ g, Eq. (49) imposes even stricter limits on the mas of the test particle. 

Now, let us consider the case of $R \gg r_{S}$.  The most accurate available information needed to evaluate the energy splitting from Eq. (47) comes from studies of the 21 cm - line of neutral hydrogen in our galaxy [19, 20].  From the wide spread of the data in the galactic latitude, Muller and Oort [20] deduce that the detected hydrogen cloud must be relatively close to the Earth.  Hence, we use in Eq. (47) a value of $R$ given by the sun-galactic center distance $R_{\odot} \approx$ 3 $\times$ 10 $^{22}$ cm [7].  For $v_{o}$, we use the sun-orbital velocity $v_{o} \approx$ 2.2 $\times$ 10 $^{7}$ cm/sec [21]. With these values, Eq. (47) yields an energy splitting $\delta E _{n} \approx$ 4.6 $\times$ 10 $^{-31}$ eV.  This result should be compared with the width of the hydrogen emission line [19] that has been measured to be about 80 kHz (which corresponds to an energy width of $\delta E \approx$ 3 $\times$ 10$^{-7}$ eV).  Obviously, the energy quantization predicted by Eq. (47) is too small to be detected via studies of the 21 cm-line for hydrogen clouds orbiting the supermassive black hole at the center of our Milky Way galaxy.

On the other hand, for the PBH with mass equal to that of the Earth, Eq.(30) yields an energy splitting $\delta E_{10} \simeq$ 2.8 $\times$ 10 $^{-6}$ eV that is one order of magnitude larger than the width of the hydrogen emission line leading to possible observable effects.

\appendix
\section{Unruh effect for orbiting test particle}
\setcounter{equation}{0}

According to Unruh [23], an accelerated mass experiences itself to be imbedded in a hot gas of photons with a temperature $T_{U}$ given by

$$
k_{B} T_{U} = \frac{a \hbar}{2 \pi c}
\eqno(A1)$$
where $a$ is the proper acceleration.  We now consider a test particle orbiting the black hole and calculate $a$ for $R$ given by the smallest stable radius $R = 3 r_{S}$.  For a circular orbit, $a$ is given by

$$
a = a_{o} \dot{t}
\eqno(A2)$$
where $\dot{t} = dt/d\tau$ is obtained from Eqs (7) and (26) in the form

$$
\dot{t} = \frac{\tilde {E}}{c^{2}(1 - r_{S}/R)} = \frac{1}{(1 - \frac{3}{2}r_{S}/R)^ \frac{1}{2}}
\eqno(A3)$$

The "ordinary" acceleration $a_{o}$ is given by

$$
a_{o} = \frac{\tilde{L}^{2}}{R^{3}}
\eqno(A4)$$

Using Eq. (16), we can express the quantity $\tilde{L}^{2}$ in terms of the radius of the orbit as follows

$$
\tilde{L}^{2} = \frac{c^{2}R^{2}}{\frac{2R}{r_{S}}- 3}
\eqno(A5)$$
Introducing this result into Eq. (A5), we have

$$
a_{o} (R)= \frac{c^{2}}{R \Big(\frac{2R}{r_{S}} - 3 \Big)}
\eqno(A6)$$

Eqs. (A2), (A3), and (A6) yield the proper acceleration for the orbiting test particle in the form

$$
a (R) = \frac{c^{2} r_{S}} {2 R^{2} \Big ( 1 - \frac{3}{2} r_{S}/R \Big)^{\frac{3}{2}}}
\eqno(A7)$$

In distinction from Ref. [23] where $a (R)$ diverges as $R \rightarrow r _{S}$, Eq. (A7) diverges as $R \rightarrow 3 r_{S}/2$.  This is not surprising, since we are considering an orbiting test particle as compared with the "standing still" particle in Ref. [23].

For $R = 3 r_{S}$, Eq. (A 7) yields $a = \frac{2 ^\frac{1}{2}c^{2}}{9 r_{S}}$.  Using this result in Eq. (A1) and invoking Eq. (38) we obtain

$$
k_{B} T_{U} = \frac{2^{\frac{3}{2}}}{9} \Big( \frac{\hbar c}{4 \pi r_{S}} \Big)\simeq 0.31 k_{B} T_{H}
\eqno(A8)$$

This result and Eq. (30) suggest that the excitation energy $\delta E_{10}$ and the proper acceleration $a (R)$ for $R = 3 r_{S}$ are related quantities.  To see this, we first express $\delta E_{10}$ in terms of the orbital period $\tau_{o}$ as follows
$$
\delta E_{10} = \frac{2 \pi \hbar}{\tau_{o}} \approx 0.136 \frac{\hbar}{c} \Big( \frac{c^{2}}{r_{S}} \Big)
\eqno(A9)$$
where Eq. (34) has been used on the right hand side.  This result is in good agreement with Eq. (30).  Consequently, we may regard Eqs. (33) and (34) as suitable starting point for further discussion of proper acceleration.

From Eq. (A7), we obtain $a \simeq 0.157 c^{2}/r_{S}$.  Using this result and Eq. (A9), we have

$$
a \simeq 1.15 \delta E_{10} \frac{c}{\hbar}
\eqno(A10)$$
Substituting this result into the Unruh formula (A1), we obtain

$$
\delta E_{10} \simeq \frac{2 \pi}{1.15} k_{B} T_{U} \simeq 1.64 {}k_{B} T_{H}
\eqno(A11)$$
where Eq. (A8) has been used to obtain the second equality.  If we use Eq. (38) for $k_{B} T_{H}$, the right hand side of Eq. (A11) becomes equal to $\frac{1.64}{4 \pi} \frac{\hbar}{r_{S}} \simeq 0. 13 \frac{\hbar}{r_{S}}$ in good agreement with Eq. (30).  This confirms the idea that the relation between $\delta E_{10}$ and the Hawking temperature $T_{H}$ discussed in Sec. 4 is actually the Unruh effect in disguise.

% Set the ending of a LaTeX document

\begin{thebibliography}{17}

\bibitem{1}
P.~A.~M. Dirac,
\emph{}
Roy. Soc. London \textbf{Ser. A133}, 60 (1931).

\bibitem{2}
A. Zee,
\emph{}
Phys. Rev. Lett. \textbf{55}, 2379 (1985).

\bibitem{3}
I. Ciufolini and A.~J. Wheeler,
\emph {Gravitation and Inertia},
(Princeton University Press, Princeton, 1995).


\bibitem{4}
W. Wilson, 
\emph{}
Phil. Mag. \textbf{29}, 795 (1915).

\bibitem{5}
A. Sommerfeld, 
\emph{}
Ann. Physik \textbf{51}, 1 (1916).

\bibitem{6}
S.~W. Hawking,
\emph{}
Nature \textbf{248}, 30 (1974); \emph{}Commun. Math. Phys. \textbf{43}, 199 (1975).



\bibitem{7}
C.~W. Misner, K.~S. Thorne, and J.~A. Wheeler,
\emph {Gravitation},
(San Francisco: Freeman, 1973).

\bibitem{8}
R. Wald,
\emph {General Relativity},
(Chicago: University of Chicago Press, 1984).


\bibitem{9}
S. Coleman, J. Preskill, and F. Wilczek,
\emph{}
Nucl.Phys. B \textbf{378}, (1992)

\bibitem{10}
K. Becker, M. Becker, and J.~H. Schwarz,
\emph {String Theory and M-Theory},
(Cambridge University Press, 2007).

\bibitem{11}
E. \v{S}im\'{a}nek,
\emph {Inhomogeneous Superconductors: Granular and Quantum Effects},
(Oxford University Press, 1994).

\bibitem{12}
L.~D. Landau and E. M. Lifshitz,
\emph {Statistical Physics},
(Pergamon Press, Oxford, 1980).

\bibitem{13}
P.~J.~E. Peebles,
\emph {Principles of Physical Cosmology},
(Princeton University Press, Princeton, NJ, 1993).

\bibitem{14}
Y.~B. Zeldovich and I.~D. Novikov,
\emph {}
Sov. Astron. \textbf {10}, 602 (1966).

\bibitem{15}
S.~W. Hawking,
\emph {}
Mon. Not. R. Astron. Soc. \textbf {152}, 75 (1971).

\bibitem{16}
A.~M. Ghez, et al,
\emph {}
Astrophysical Journal \textbf {689}, 1044 (2008).

\bibitem{17}
G. Baym,
\emph {Lectures on Quantum Mechanics},
(W. A. Benjamin, Inc, New York, Amsterdam, 1969).

\bibitem{18}
D.~N. Page,
\emph {}
Phys. Rev.  \textbf {D14}, 3260 (1976).

\bibitem{19}
H.~I. Ewan and E.~M. Purcell,
\emph {}
Nature \textbf {168}, 4270 (1951).

\bibitem{20}
C.~A. Muller and J.~H. Oort,
\emph {}
Nature \textbf {168}, 4270 (1951).

\bibitem{21}
E. Chaisson and S.~McMillan,
\emph {Astronomy Today},
(Prentice Hall, New Jersey, 1993).

\bibitem{22}
W.~G. Unruh,
\emph {}
Phys. Rev. \textbf {D14}, 870 (1976).














\end{thebibliography}
\end{document}